%
\let\useblackboard=\iftrue
%
%
\newfam\black
\input harvmac.tex
%
\ifx\epsfbox\UnDeFiNeD\message{(NO epsf.tex, FIGURES WILL BE
IGNORED)}
\def\figin#1{\vskip2in}
\else\message{(FIGURES WILL BE INCLUDED)}\def\figin#1{#1}\fi
\def\ifig#1#2#3{\xdef#1{fig.~\the\figno}
\midinsert{\centerline{\figin{#3}}%
\smallskip\centerline{\vbox{\baselineskip12pt
\advance\hsize by -1truein\noindent{\bf Fig.~\the\figno:} #2}}
\bigskip}\endinsert\global\advance\figno by1}
\noblackbox
\def\Title#1#2{\rightline{#1}
\ifx\answ\bigans\nopagenumbers\pageno0\vskip1in%
\baselineskip 15pt plus 1pt minus 1pt
\else
\def\listrefs{\footatend\vskip
1in\immediate\closeout\rfile\writestoppt
\baselineskip=14pt\centerline{{\bf
References}}\bigskip{\frenchspacing%
\parindent=20pt\escapechar=` \input
refs.tmp\vfill\eject}\nonfrenchspacing}
\pageno1\vskip.8in\fi \centerline{\titlefont #2}\vskip .5in}
 
scaled\magstep3
 
scaled\magstep3
 
scaled\magstep3
 
scaled\magstep3
 
scaled\magstep3
\ifx\answ\bigans\def\tcbreak#1{}\else\def\tcbreak#1{\cr&{#1}}\fi

\useblackboard
\message{If you do not have msbm (blackboard bold) fonts,}
\message{change the option at the top of the tex file.}

\font\blackboard=msbm10 scaled \magstep1
\font\blackboards=msbm7
\font\blackboardss=msbm5
\textfont\black=\blackboard
\scriptfont\black=\blackboards
\scriptscriptfont\black=\blackboardss

\else

\fi
%

%
\def\yboxit#1#2{\vbox{\hrule height #1 \hbox{\vrule width #1
\vbox{#2}\vrule width #1 }\hrule height #1 }}
\def\fillbox#1{\hbox to #1{\vbox to #1{\vfil}\hfil}}
\def\ybox{{\lower 1.3pt \yboxit{0.4pt}{\fillbox{8pt}}\hskip-0.2pt}}

\def\comments#1{}

\def\half{{1\over 2}}
\def\Tr{{{\rm Tr\  }}}

\def\II{\relax{I\kern-.07em I}}

\def\inbar{\,\vrule height1.5ex width.4pt depth0pt}
\def\IZ{\relax\ifmmode\mathchoice
{\hbox{\cmss Z\kern-.4em Z}}{\hbox{\cmss Z\kern-.4em Z}}
{\lower.9pt\hbox{\cmsss Z\kern-.4em Z}}
{\lower1.2pt\hbox{\cmsss Z\kern-.4em Z}}\else{\cmss Z\kern-.4em
Z}\fi}
\def\IB{\relax{\rm I\kern-.18em B}}
\def\IC{{\relax\hbox{$\inbar\kern-.3em{\rm C}$}}}
\def\ID{\relax{\rm I\kern-.18em D}}
\def\IE{\relax{\rm I\kern-.18em E}}
\def\IF{\relax{\rm I\kern-.18em F}}
\def\IG{\relax\hbox{$\inbar\kern-.3em{\rm G}$}}
\def\IGa{\relax\hbox{${\rm I}\kern-.18em\Gamma$}}
\def\IH{\relax{\rm I\kern-.18em H}}
\def\IK{\relax{\rm I\kern-.18em K}}
\def\IP{\relax{\rm I\kern-.18em P}}
\def\pp{{\relax{=\kern-.42em |\kern+.2em}}}

\font\cmss=cmss10 \font\cmsss=cmss10 at 7pt
\def\IR{\relax{\rm I\kern-.18em R}}

\def\Tr{{\rm Tr\ }}

\def\frac#1#2{{{#1} \over {#2}}}

%
%

\def\NP{{\it Nucl. Phys.\ }}
\def\NPPS{{\it Nucl. Phys. Proc. Suppl.\ }}

\def\PL{{\it Phys. Lett.\ }}
\def\PR{{\it Phys. Rev.\ }}
\def\PRL{{\it Phys. Rev. Lett.\ }}
\def\CMP{{\it Comm. Math. Phys.\ }}

\def\JHEP{{\it JHEP \ }}
\def\ATMP{{\it ATMP \ }}
\writedefs

\Title{ \vbox{\baselineskip12pt\hbox{hep-th/9810075}
\hbox{BROWN-HET-1145}
}}
{\vbox{
\centerline{Constraints on Higher Derivative Operators}
\centerline{in the Matrix Theory Effective Lagrangian}}}

\centerline{ David A. Lowe}
\medskip

\centerline{Department of Physics}
\centerline{Brown University}
\centerline{Providence, RI 02912, USA}
\centerline{\tt lowe@het.brown.edu}
\bigskip

\centerline{\bf{Abstract}}

The consistency of Matrix theory with supergravity requires that in
the large $N_c$ limit terms of 
order $v^4$ in the $SU(N_c)$ Matrix effective potential are not
renormalized beyond one loop in perturbation theory. For $SU(2)$ gauge
group, the required   
non-renormalization theorem was proven recently by Paban, Sethi and
Stern. In this paper we consider the constraints supersymmetry imposes 
on these terms for groups $SU(N_c)$ with $N_c>2$. 
Non-renormalization theorems are proven for certain tensor
structures, including the structures that appear in the one-loop
effective action. However it is expected other tensor structures can
in general be present, which may suffer renormalization at three loops
and beyond.

\vfill
\Date{\vbox{\hbox{\sl October, 1998}}}

\lref\wati{W. Taylor, ``Lectures on D-branes, Gauge Theory and
M(atrices),'' hep-th/9801182.}
\lref\bfss{T. Banks, W. Fischler, S.H. Shenker and L. Susskind, ``M
Theory as a Matrix Model: A Conjecture,'' \PR {\bf D55} (1997) 5112, 
hep-th/9610043.}
\lref\seiberg{A. Sen, ``D0-Branes On T**N And Matrix Theory,''
\ATMP {\bf 2} (1998) 51, hep-th/9709220;
N. Seiberg, ``Why Is The Matrix Model Correct?''
\PRL {\bf 79} (1997) 3577, hep-th/9710009.}
\lref\yoneya{Y. Okawa, T. Yoneya,
``Multibody Interactions Of D Particles In Supergravity And
Matrix Theory,'' hep-th/9806108.}
\lref\pouliot{J. Polchinski and P. Pouliot, ``Membrane Scattering With
M Momentum Transfer,'' \PR {\bf D56} (1997) 6601, hep-th/9704029.}
\lref\sethi{S. Sethi and Mark Stern, ``D-Brane Bound States Redux,''
\CMP {\bf 194} (1998) 675, hep-th/9705046.}
\lref\vijay{V. Balasubramanian, R. Gopakumar and F. Larsen, ``Gauge
Theory, Geometry and the Large N Limit,'' \NP {\bf B526} (1998) 415,
 hep-th/9712077.}
\lref\kabat{D. Kabat, W. Taylor, ``Spherical Membranes in Matrix
Theory,'' \ATMP {\bf 2} (1998) 181, 
hep-th/9711078; ``Linearized Supergravity From Matrix
Theory,'' \PL {\bf B426} (1998) 297, hep-th/9712185.}
\lref\ramsd{W. Taylor and M. Van Raamsdonk,``Angular 
Momentum And Long Range Gravitational Interactions In
Matrix Theory,'' hep-th/9712159.}
\lref\becker{K. Becker and M. Becker, ``A Two Loop Test Of M(atrix)
Theory,'' \NP {\bf B506} (1997) 48, hep-th/9705091; K. Becker,
M. Becker, J. Polchinski and A. Tseytlin,  
``Higher Order Graviton Scattering In M(atrix) Theory,'' \PR {\bf D56} 
(1997) 3174, hep-th/9706072.}
\lref\malda{J. Maldacena, ``The Large N Limit of Superconformal Field
Theories and Supergravity,'' hep-th/9711200.}
\lref\paban{S. Paban, S. Sethi and M. Stern, ``Constraints From
Extended Supersymmetry in Quantum Mechanics,'' hep-th/9805018.}
\lref\pabant{S. Paban, S. Sethi and M. Stern, ``Supersymmetry and
Higher Derivative Terms in the Effective Action of Yang-Mills
Theories,''
\JHEP {\bf 06} (1998) 012, hep-th/9806028.}
\lref\pabantt{S. Paban, S. Sethi and M. Stern, ``Summing Up Instantons
in Three-Dimensional Yang-Mills Theories,'' hep-th/9808119.}
\lref\lowe{D.A. Lowe, ``Statistical Origin of Black Hole Entropy in
Matrix Theory,'' \PRL {\bf 81} (1998) 256, hep-th/9802173.}
\lref\douglas{M.R. Douglas, D. Kabat, P. Pouliot and S. Shenker,
``D-Branes and Short Distances in String Theory,'' \NP {\bf B485}
(1997) 85, hep-th/9608024.}
\lref\malprob{J. Maldacena, ``Probing near-extremal black holes with
D-branes,'' \PR {\bf D57} (1998) 3736, hep-th/9705053; 
``Branes Probing Black Holes,''
\NPPS {\bf 68} (1998) 17, hep-th/9709099.}
\lref\esko{E. Keski-Vakkuri and P. Kraus,``Notes on Branes in Matrix
Theory,'' hep-th/9706196; ``Born-Infeld Actions from Matrix Theory,''
hep-th/9709122.}
\lref\tseytlin{I. Chepelev and A. Tseytlin, ``Long-distance
interactions of branes: correspondence between supergravity and super
Yang-Mills descriptions,'' \NP {\bf B515} (1998) 73, hep-th/9709087.}
\lref\dine{M. Dine, R. Echols and J.P. Gray, ``Tree Level Supergravity 
and the Matrix Model,'' hep-th/9810021.}
\lref\lowelor{D.A. Lowe, ``Eleven-Dimensional Lorentz Symmetry from
SUSY Quantum Mechanics,'' \JHEP {\bf 10} (1998) 003, hep-th/9807229.}
\lref\wilson{K.G. Wilson, T. Walhout, A. Harindranath, W.M. Zhang,
R.J. Perry and S. Glazek, ``Nonperturbative Light-Front QCD,'' \PR
{\bf D49} (1994) 6720, hep-th/9401153.}

\newsec{Introduction}

One of the exciting themes to emerge from recent work is the duality
between
gauge theories and gravitational theories \refs{\bfss,
\malda}. A prime example is M-theory in discrete light-cone gauge
which is thought to be
described by the supersymmetric quantum mechanics of low-energy
D-particles \refs{\seiberg}. Another example is supergravity in
$d$-dimensional
anti-de Sitter space which is conjectured to be described by
$d-1$-dimensional 
large $N_c$ superconformal field theory with $SU(N_c)$ gauge
group \malda.

In the case of M-theory, we are often interested in an effective action 
expanded in powers of velocities \refs{\becker \douglas \malprob 
\vijay \tseytlin \ramsd \kabat {--} \yoneya}. 
The effective action computed using the
quantum mechanics can be compared straightforwardly with the
supergravity results. For the Matrix theory results to be consistent
with supergravity, agreement is required in the large
$N_c$ limit. This is one of the 
key assumptions that enters into the statistical 
derivation of black hole entropy from Matrix theory \lowe.
Although exact agreement is required only in the 
large $N_c$ limit, some remarkable results have been obtained
for the leading order terms in the expansion for the $SU(2)$ case.
In particular, for the $SU(2)$ quantum mechanics,
non-renormalization theorems have been proven for the $v^4$ and $v^6$
terms \refs{\paban, \pabant}. 

In this paper we consider the constraints supersymmetry imposes on the 
$v^4$ terms for the general $SU(N_c)$ case. It is shown certain tensor 
structures are not renormalized beyond one loop. However, for more
general tensor structures, it is argued
renormalization should be expected at order $v^4$ for $N_c>2$, which
may begin to appear at three-loop order and beyond.

\newsec{Matrix review}

The Lagrangian of Matrix theory is
\eqn\mlag{
L = {1\over g^2} \Tr \bigl( (D_0 X^i)^2 + \half [X_i,X_j]^2 + 
i \psi_a D_0 \psi_a - \psi_a \gamma^i_{ab} [X_i, \psi_b] \bigr)~,
}
where all the fields are in the adjoint of $SU(N_c)$. The theory is
supersymmetric with respect to $16$ supercharges.

We will be interested in the effective Lagrangian of this theory 
at a point where the $X$'s are diagonal, expanded in a 
power series in velocity. We will use the basis $x_A$ to denote the
elements of the Cartan subalgebra, with $A=1,\cdots, N_c-1$. The $x^A$ 
correspond to the $N_c-1$ relative displacements of $N_c$ D-particles.
The effective Lagrangian schematically takes the form
\eqn\efflag{
L_{eff} = {1\over g^2} \sum_n v^{2n} f_n(r)~,
}
where perturbatively the $f_n$ are
\eqn\pcoeff{
f_n(r) = \biggl( {1\over r^4}\biggr)^{n-1} \sum_l C_{nl} 
\biggl( {g^2 \over r^3}\biggr)^l~,
}
and $l$ is the number of loops \becker.

The terms of order $v^4$ will be the focus of this paper. At one loop, 
the purely bosonic terms take the simple form
\eqn\voneloop{
L_1 = {15\over 16}\biggl( \sum_A {v_A^{4} \over x_A^{7}} + 
\sum_{A<B} { (v_A -v_B)^4 \over (x_A-x_B)^7} \biggr)~.
}
This agrees with the formula obtained from linearized supergravity,
using the graviton 
propagator corresponding to zero longitudinal momentum transfer.

\newsec{Constraints from supersymmetry}

The object of this paper is to analyze the constraints supersymmetry
imposes on the terms of order $v^4$ in the effective action.
The general strategy for the analysis of will follow that of \paban,
where it was found that supersymmetry ensures \voneloop\ receives no
corrections for the $SU(2)$ case, even at the non-perturbative level.
We consider therefore the 
supersymmetric variation of the eight fermion terms that 
arise in the supersymmetric completion of the $v^4$ term. 
Part of this variation will be a nine fermion term that 
cannot be canceled by any other source.
Demanding this nine fermion term vanish gives strong constraints on the 
general form of the eight fermion term that appears in the effective
action.

The supersymmetry transformations take the form
\eqn\susytrans{
\eqalign{
\delta x_A^{i} &= -i \epsilon \gamma^i \psi_A + \epsilon N_{AB}^i \psi_B \cr
\delta \psi_{aA} &= (\gamma^i v_A^{i} \epsilon)_a + (M_A \epsilon)_a ~,\cr}
}
where $i$ 
labels the vector of $Spin(9)$ and $a$ the $16$ component spinor.
The effective Lagrangian is expanded in powers of $n$, which 
counts the number of time derivatives plus half the number of
fermions. 
$N$ and $M$ encode the higher order corrections to the supersymmetry 
transformations.

The first step in the calculation is to show that the metric
multiplying 
the $v^2$ term in the Lagrangian is necessarily flat. $N$ must vanish
at leading order ($n=1$) since no gauge invariant $Spin(9)$ symmetric
terms can be constructed. At leading order, $M$ can be order $n=2$ and 
we must check this term vanishes. Let us consider the closure of the
supersymmetry algebra
\eqn\closure{
[ \delta_{\epsilon_1} , \delta_{\epsilon_2} ] x_A^{i} = - 2 i \epsilon_2
\epsilon_1 v_A^{i} - i \epsilon_{2a} ( \gamma^i M_{abA} +
M_{baA}\gamma^i) \epsilon_{1b}~.
}
The last term must vanish for the algebra to close. For 
matrices with $Spin(9)$ spinor indices, this leads to the condition
that
$M^A=0$ for all $A$ at this order. This implies the metric is flat, so 
corrections to $M$ can begin at order $n=3$ and corrections to 
$N$ should begin at order $n=2$.

Next we must consider the general form of the eight fermion terms that 
can appear in the effective action. They are built out of the general 
fermion bilinears
\eqn\fbilin{
\psi^A \Gamma \psi^B~,
}
where $\Gamma \in \{ I, \gamma^i, \gamma^{ij}, \gamma^{ijk},
\gamma^{ijkl} \}$.  The $\gamma^i$ represent the $Spin(9)$ Clifford 
algebra and the other matrices are defined as
\eqn\gammats{
\eqalign{
\gamma^{ij} &= {1\over 2!} [\gamma^i, \gamma^j] \cr
\gamma^{ijk} &= {1\over 3!} (\gamma^i \gamma^j \gamma^k - \gamma^j
\gamma^i \gamma^k + \cdots)  \cr
\gamma^{ijkl} &= {1\over 4!} (\gamma^i \gamma^j \gamma^k \gamma^l -
\gamma^j \gamma^i \gamma^k \gamma^l + \cdots )~. \cr}
}
The Lorentz indices of a general product of four of the fermion 
bilinears are contracted with a product of between zero and sixteen 
$x^{iA}$. Finally the gauge indices are contracted with a tensor
function of the $x^{iA}$, that respects $Spin(9)$ invariance. 
The sum of terms must respect 
the residual Weyl invariance of the underlying $SU(N_c)$ gauge
theory. 
If we represent the $x_A$ as $x_A = e_{A} -e_{A+1}$, the Weyl group acts by
permutations on the set of $N$ objects $e^A$ and $e^{N}$.
For the tensor functions to agree with perturbation theory we 
know they must fall off in any direction as the $x$'s go to infinity.

The nine fermion term produced in the supersymmetric variation of the 
eight fermion term $f^{(8)} \psi^8$ is
\eqn\nineferm{
 \gamma^n_{ab} \psi_{aA} {\partial \over \partial x_A^{n} }
(f^{(8)} \psi^8) =0~.
}
This gives us a set of coupled partial differential equations to solve 
for the general supersymmetry constraints. 

As we will see in a moment, it is convenient to derive a weaker set of 
second order partial differential equations from \nineferm. Act on
\nineferm\ with the operator $\gamma^n_{bc} {d\over d\psi_{cM}}{\partial 
\over \partial x_N^n}$ to give the equation
\eqn\prelapl{
16 {\partial^2 \over \partial x_M^n \partial x_N^n} (f^{(8)} \psi^8 ) -
{\partial^2 \over \partial x_L^n \partial x_N^n} \psi_{aL} ( f^{(8)} 
{d \over d \psi_{aM} } \psi^8)
- {\partial^2 \over \partial x_L^m \partial x_N^n} \psi_{aL}
\gamma^{mn}_{ac}( f^{(8)} 
{d \over d \psi_{cM} } \psi^8)
=0~.
}

Let us recall how the solution of these equations works for the 
$SU(2)$ case \paban. The most general
eight fermion term is
\eqn\eigtfers{
\psi \gamma^{ij} \psi \psi \gamma^{jk} \psi \psi
\gamma^{lm} \psi \psi \gamma^{mn} \psi \biggl( g_1(r) \delta_{in}
\delta_{kl} + g_2(r) \delta_{kl} x_i x_n + g_3(r) x_i x_k x_l x_n\biggr)~.
}
For the $SU(2)$ case, \prelapl\ simply reduces to the Laplacian.
It is easiest to
start by considering the action of the Laplacian on the last term in
\eigtfers. By considering the independent tensor structures that arise 
after acting with the Laplacian, a decoupled equation is obtained for
the unknown function $g_3$. The solution is a power law
$g_3 = c/r^{15}$, with
$c$ an undetermined constant. The
equations for the other functions are coupled to $g_3$. 
Solving these equations
yields $g_2 = -4 c/13 r^{13}$ and $g_1 = 2c/143 r^{11}$. The other
integration constants that arise give solutions with lower inverse
powers of $r$. These terms must vanish for the solution to agree with
perturbation theory in the limit $g\to 0$. The only physically
consistent solution has scaling appropriate to the one-loop effective
action, therefore these terms must not be renormalized.

The key fact that yields the non-renormalization
theorem in the $SU(2)$ case is that a unique term, with the
maximal number of $x$'s contracted with fermion bilinears, satisfies an
equation that decouples from the other unknown functions. The only
solution to this equation has a definite scaling with respect to $r$
which corresponds to one-loop behavior. All the terms with lower
number of $x$'s are determined in terms of this term, and the
physical boundary conditions.

We would like to generalize this statement to $SU(N_c)$ with
$N_c>2$. Unfortunately, the equations \prelapl\ in general lead to a
complicated system of linear second-order partial differential
equations for which the general solution is difficult to construct. 
Before commenting further on the general case let us consider a
special set of tensor structures that may be analyzed explicitly.

These correspond to terms that have a nontrivial limit when some of
the fermions are constrained to be equal and the rest vanish.
To see why we expect a simplification in
this case 
consider first $N_c-1$ D-particles moving with the same
velocity, which may then be Galilean boosted to zero. The positions of 
the particles remain generic in this limit. A probe
D-particle with some finite velocity $v$ 
now sees a supersymmetric system, so its
worldline action is tightly constrained and we can
expect a non-renormalization theorem. 
The analogous statement for the eight fermion term will be that if a
term has a nontrivial limit when the $\psi^A$ for $A>1$ are 
constrained to vanish, the unknown function of the Casimirs 
multiplying it should be
determined.

This statement may be generalized 
straightforwardly to the case when $N_c-N_0$ D-particles remain at rest
and are probed by a collection of $N_0$ D-particles all moving with the
same velocity $v$. Rather than taking the non-vanishing $\psi^A$ to
lie in a regular $SU(2)$ subgroup of $SU(N_c)$ as above, we simply
take the single independent non-vanishing $\psi$ to lie in an
non-regular $SU(2)$ subgroup of $SU(N_c)$.

Let us consider then the system of equations \prelapl\ for the case
$M=1$, $N=1$ after setting all $\psi^A=0$ for $A>1$. The result is 
\eqn\lapl{
{\partial^2 \over \partial x_n^1 \partial x_n^1} (\psi_1^8 f^{(8)} ) =0~.
}
Again we will first focus on terms with the maximal 
number (four) of $x^1_j$'s contracted with the fermion
bilinears. Once
again we obtain a decoupled equation for the unknown function $g_3(x^A)$ of
the $x$'s appearing in this term, which takes the same form as for the
$SU(2)$ case. Now, however, the solution of the
Laplacian can be singular not just at $x^1=0$, but at any point where
the corresponding D-particles collide. The solutions should fall off
in the limit that $x_1\to \infty$. The
general solution then is 
\eqn\gensol{
g_3 = {c_0 \over x_1^{15}} +\sum_{A>1} {c_A \over (x_1 -x_A)^{15}} ~,
}
where $c_0$ and $c_A$ are functions of $x_B$, for $B>1$. The $c$'s can be
determined by considering the limit that $x_1\to x_B$ for some $B$ or
$x^1\to 0$,
when $SU(2)$ gauge symmetry is restored. To agree with the $SU(2)$
result \paban, the $c$'s must be independent of the $x_A$. Since the
$c$'s are constant, we see these terms have scaling with $x$ such that 
they can only be generated at one loop. Likewise, one can make a
similar argument for terms with fewer $x_1$'s contracted with fermion
bilinears. These are determined using the boundary conditions and the
solution for $g_3$ above.
This proves all the terms with a nontrivial limit when $N_c-2$ of the $\psi_A$
vanish are not renormalized beyond one loop. The same argument goes
through unchanged
when the non-vanishing $\psi$ lies in an non-regular
$SU(2)$ subgroup of $SU(N_c)$.
Assuming these
eight fermion terms uniquely determine the bosonic terms (a result
that is expected, but thus far not rigorously proven, even for the
$SU(2)$ case) \voneloop\ contains all the purely bosonic terms with a 
nontrivial limit when only one independent velocity parameter appears.

However, more generally there can be terms that vanish when all but one 
of the
$\psi_A$ are zero. 
In this case it does not appear that the equations \prelapl\
decouple
and allow the solutions to be classified directly. 
Rather than analyze the second order equations \prelapl, it is perhaps 
more efficient to
consider the stronger set of first order equations obtained 
by acting with $\gamma^i_{ac} x_i^M d/d\psi^N_c$. However these
equations likewise do not decouple in any obvious way to allow the
solutions to be classified explicitly.

\newsec{Discussion}

It is possible then that supersymmetry and gauge invariance alone are not 
sufficient to uniquely fix all 
the order $v^4$ terms in the effective action for
$N_c>2$. 
For finite $N_c$ we do not expect any additional symmetries to be 
relevant, so assuming the previous statement is valid, 
renormalization of these terms can be expected. For large
$N_c$
additional symmetries may appear (for example eleven-dimensional 
Lorentz symmetry as discussed in \lowelor). It is possible these
symmetries are sufficient to restrict the form of the solutions so
that only terms of the one-loop form can appear.

A heuristic argument why in general renormalization should be expected 
can be
made as follows. Consider a background of $N_c-1$ D-particles with generic
velocities. If another D-particle is used to probe this background,
the worldline action for this particle will not be supersymmetric, 
so no general non-renormalization theorem would be expected to hold. 
It seems likely even the non-renormalization theorem at
order $v^4$ will fail in this case. Terms in the worldline action
that are not renormalized
will be mapped by supersymmetry into terms that must vanish. On the
other hand, terms that can be renormalized will be mapped by
supersymmetry to terms that at best vanish after summing over a number 
of contributions. This appears to be the structure of the differential 
equations discussed above. For the general case the equations do not
appear to decouple and lead to a unique solution.

No renormalizations of $v^4$ terms are present at
two-loop order \yoneya\ for planar diagrams. We have also checked the
non-planar diagrams vanish in this case. Such terms may arise at
three-loop order and beyond. It should be noted that if such terms
appear in perturbation theory they will be accompanied by powers of
$N_c$ that are not subleading at large $N_c$. This will spoil the
agreement between perturbative Matrix theory and low-energy supergravity.

Related terms have been considered in 
\dine\ at order $v^6$ and three loops 
which apparently renormalize the two-loop $v^6$ answer of
\yoneya, for $SU(N_c)$ with $N_c>4$, and lead to disagreement with the 
supergravity result. In contrast, for
 $SU(2)$, the entire set of $v^6$ terms satisfies a non-renormalization
theorem \pabant. If these renormalizations appear for $v^6$ at
three loops, one would expect them to contribute to the imaginary part 
of higher loop $v^4$ amplitudes via the unitarity relations.

A disagreement between low-energy supergravity and
Matrix theory for sufficiently generic amplitudes may simply be a sign 
that higher-order 
counter-terms must be added to the Matrix quantum mechanics to regain
eleven-dimensional Lorentz invariance. Of course this is the standard
state of affairs in the light-front quantization of quantum 
chromodynamics \wilson, where counter-terms are fixed by the
requirement of Lorentz invariance. The only novelty in Matrix theory
may be that
these counter-terms are not needed for certain low-order calculations
due to the high degree of supersymmetry. If this is the case, it calls 
into question the use
of Matrix theory as a fundamental definition of M-theory.

\bigskip
{\bf Acknowledgments}

I thank A. Jevicki, A. Lawrence, S. Ramgoolam, S. Sethi and T. Yoneya
for helpful discussions, and Mark Stern for comments on an
earlier version of the manuscript.
I thank the Aspen Center for Physics for hospitality during
the course of this research.
This research is supported in part by DOE grant DE-FE0291ER40688-Task A.

\listrefs
\end